\documentclass[final,1p,10pt]{elsarticle}

\usepackage{graphicx}
\usepackage{amsmath}
\usepackage{amsfonts}
\usepackage{amssymb}

\journal{Physica A: Statistical Mechanics and its Applications}

\begin{document}

\begin{frontmatter}

\title{Effects of Publications in Proceedings  on the Measure of the Core Size of Coauthors}

\author[lab1,lab2]{Janusz Mi\'{s}kiewicz}
 \address[lab1]{ Institute of Theoretical Physics, University of Wroc\l{}aw, pl. M. Borna 9, 50-204 Wroc\l{}aw, Poland}
 \address[lab2]{Department of Physics and Biophysics, Wroc\l{}aw University of Environmental and Life Sciences. ul. Norwida 25, 50-375 Wroc\l{}aw}
\ead{jamis@ift.uni.wroc.pl}

\begin{abstract}
Coauthors (CA) of a "lead investigator" (LI) can receive a rank ($r$) according to their "importance" in having published joint publications with the LI. It is commonly accepted, without any proof, that  publications in peer review journals and e.g. conference proceedings do not have the same "value" in a CV.  Same for papers contributed to encyclopedia and book chapters. It is here examined whether the relationship between the number ($J$) of  publications of  some scientist with her/his coauthors, ranked according to their decreasing importance, i.e.  $ J \propto 1/r^{\alpha} $, as found by Ausloos \cite{Sofia3a}, still holds if the overall publication list is broken into such specific types of publications.  Several authors, with different careers, but mainly having worked in the field of statistical mechanics, are studied here  to sort out  answers to the questions. The exponent $\alpha$ turns out to be weakly scientist dependent, only  if the maximum value of $J$ and $r$ is large and is $\sim +1$ 
then. The $m_A$ core value, i.e. the core number of CAs,  for proceedings only is about half of the total one, i.e.  when all publications are counted. Contributions to the numerical values from  both  encyclopedia and book chapters are marginal. The role of a time span on $m_A$  is also examined in  two cases in relation to career  activity considerations. It can considered that the findings serve as a contrasting point of view on how to quantify an individual (publication) career as recently done by Petersen et al.  \cite{HESPRE2010career,HESPNAS2012career,HESPNAS2011career}, here emphasizing  the collaboration size and   evolution,  rather than a citation count, moreover specifying the type of publication. Through the various $m_A$'s  one can distinguish different behavior patterns of a scientist  publication with CAs.
 \end{abstract}
\end{frontmatter}

\section{Introduction  }\label{sec:intro}

In order to  justify and/or promote a young researcher  work, he/she is often sent to present  his/her research at scientific meetings, and publish the research results  in proceedings.  It is also somewhat  commonly accepted, without any proof, that proceedings papers contain more coauthors than peer review journals.  One does not question here  whether one should be considering the publications in proceedings and those in peer review journals  with an equal weight to measure the value of some scientific report. In favor of publishing contributions in proceedings, it  seems to be one way  to  justify more quickly  the time spent by a visitor in a team or  in a laboratory, - because conference proceedings  are thought to be less strict or take less time at the reviewer level than  well established peer review journals.  Necessarily, the team leader or laboratory coordinator is associated to such publications, justifying his/her principal investigator  (PI) status. Of course, the PI number of publications is 
then  increased much. One basic question  which would seem to be raised on such aspects of scientific life  concerns the number of relevant coauthors for  the set of publications of a PI. 
One may wonder about their  quantitative role for a PI or the team.

Ausloos "coauthor core" definition and its subsequent measure \cite{Sofia3a} tackle such considerations in a constructive way, through the relationship between the number ($J$) of (joint) publications with coauthors ranked according to rank ($r$) importance. The approach presents great differences  with respect to the recent Petersen et al. career life time, growth and/or decay considerations \cite{HESPRE2010career}   later  expanded in   \cite{HESPNAS2012career,HESPNAS2011career}.  These  emphasize  the citation count within some   Hirsch index idea  \cite{hindex,hirsch10}, taking into account some normalization based on the group size and   its time evolution,  yet mainly  testing  the popularity of a  $paper$.  Ausloos' approach emphasizes the role of $persons$ in a team scientific production, in relation  with a PI, rather than citations.
 
In fact, the PI   behind and in  a publication is sometimes  hard to define. It is known that the notion of PI  arises from administrative considerations. Sometimes a coauthor (CA), not necessarily receiving the first place in a coauthor list on a publication, has played an important role in the scientific investigations.  However, without a criterion, one cannot write that such a CA has done "more work" than a PI, - or conversely.  That is one of the main reasons why the wording PI will be thereafter abandoned and replaced by LI  standing for "leading investigator", e.g., as someone accepted by the community as a well known  leader for some investigation. 
  
Usually the position on a CA list in a publication hints toward some responsability, but not always. Expected  emphasis on the position of a scientist on a publication list is a delicate matter when quantifying a contribution. The more so because it is thought that  abusive coauthorship and publication parasitism exist, as emphasized by Kwok \cite{Kwok05}, i.e.  the "White Bull effect". Sometimes indeed,  a large quantity of so called proceedings papers or invited lectures have many coauthors, - usually  in order to take into account various contributions on the reviewed subject and/or to promote team size visibility, among likely many other reasons.
 
On the other hand, recall that   the   $m_A$-index  \cite{Sofia3a}  measures the core of coauthors in a research team, as if centered on a researcher, a LI, who can be anyone. The   $m_A$-index is deduced from a plot of the number ($J$) of     joint   publications of this LI  with CAs ranked according to their rank ($r$) of  importance; $r=1$ being the most prolific CA with the LI. 
 
 Ausloos \cite{Sofia3a} has found a  simple power law relating  $J$ and $r$
  \begin{equation}   
 J \propto 1/r^{\alpha}.
\label{eq1}
  \end{equation}
The  power law exponent  $\alpha$ is not exactly +1. It depends on the examined data range,  -- as  usual; this is well known \cite{SIAM51.09.661powerlaws}.   One may  also conjecture that irregularities and deviations from such a simple analytic law, Eq. (\ref{eq1}), maybe due to: publication inflation, proceedings counting,  coauthorship inflation, for whatever reason \cite{JASIST52.01.610lotkabreakHKRR}, - even considering that all counts are made on a reliable data basis.    
 Nevertheless, from 
  this point of view, one can  derive the  $m_A$-index, giving the core of coauthors range, through a condition similar to that defining the $h$-index of a scientist \cite{hindex,hirsch10}, i.e.  
\begin{equation}\label{eq2}
 m_A \; \equiv \; r,  such \; that  \; r\; \le \; J.
  \end{equation}

Most likely,  Ausloos law, Eq. (\ref{eq1}),  seems to be best for large teams, and/or for authors having many publications and many coauthors. Indeed, as pointed out  already in  \cite{Sofia3a},   when an author has not many publications, or   has few coauthors, the law might  be a statistically poor description of the (rare) empirical data. On the other hand,  deviations in presence of a large set of publications  and a large set of coauthors might be due to several causes.  

So called "intrinsic causes"  might arise from a large productivity of the group based on a high turnover of young and junior researchers, with $r>> 1$,  but having few $J$ with the LI. On the other  $r$ range, i.e., for small $r$,  many  contributions may arise  from stable  partners who provide visibility of the team by going to conferences and summer schools.  
Among "extrinsic causes",  one can simply also mention funding conditions or applications for funding constraints requesting to show the rather large size of the LI team.  Indeed, public and private research funding agencies   $claim$ to search for  and to promote such collaborations.  Again, to estimate the quality of such CA  is far from obvious, nor to measure their   internal impact  for  the team and LI. 
 
Therefore it is of interest to examine the generality of Eq. (\ref{eq1}).  This is made here by breaking  the overall publication list  into specific types of publications, like in peer reviewed journals, proceedings, invited chapters of books or encyclopedia, ... .  Several LI cases  are studied here. As examples,  two well known LI in statistical physics: (i)  H.E. Stanley (HES), the most prolific of such authors (with the highest $h$-index known for physicists,   $h>115$, though this is irrelevant for the present purpose); and (ii) D. Stauffer (DS), an accepted leader in theoretical and numerical statistical physics.
Moreover, (iii) M. Ausloos (MA), who invented the $m_A$-index   \cite{Sofia3a} is included, due to his large publication list with CAs, in the same quantitative range as DS, but  with papers much less quoted than those of  HES  or  DS. 
 
For comparison,  the publication lists by 6 scientists having worked or still working in statistical mechanics,  in Wroclaw, PL are included, though  it is  not claimed  that any is a PI, -  at the anglo-saxon or even USA  status\footnote{The present author does not even claim to be called a LI}.  Since the publication list for several of them,  being "rather short" as compared to that of HES, DS and MA,  it is emphasized that  the data is mainly given for covering a large range of $J$ and $r$.  Yet, its analysis supports Ausloos' claim that there is something else that the $h$-index for measuring  the quality or usefulness of a scientific author, i.e. his/her capacity in building a research team with relevant CAs.

The methodology, about separating and counting different types of publications,  is  briefly explained in Sect. \ref{sec:Method}. The data analysis of the  coauthorship features is reported in Sect. \ref{sec:data}, distinguishing between all publications, Sect. \ref{sec:dataanalNJP}, and those in proceedings,
Sect.  \ref{sec:dataanalNJPp}.
In Sect. \ref{sec:dataset}, some discussion on the statistical (mechanics)  aspects of these illustrative cases are presented in line with general considerations on "subcores" of  coauthors for a LI.   Moreover, it is  studied whether one should consider any time dependence of the present findings. Following the two time intervals considered by HES for his publication list, the same time intervals have been used for   "comparing" HES and DS CAs cores as a function of time interval in Sect. \ref{sec: Appendix}. Sect. \ref{sec:conclusions} serves as a conclusion with suggestion for future work and applications.  
 
\section{Methodology} \label{sec:Method}

As mentioned in the introduction, in  order to quantify Ausloos   law and verify its validity limit,  in the here above defined context, several publication lists have been selected. On one hand, the publication  list, with mentioned  coauthors of HES, DS, and MA   are available   through web sites and those of the Wroclaw scientists through personal contacts.   On the other hand, the first three are large enough,   to allow for expecting reasonable  and meaningful statistical work. Since mine (JM) and those of the Wroclaw scientists are shorter,  they can surely  serve as a measure of variability of the findings. 

In each list, many chapters in books and encyclopedia, or  papers subsequent to scientific presentation at various    scientific meetings can be  "easily" distinguished from  papers in journals with peer review,  
searching when in doubt, e.g. for "unusual  journals",  whether a conference had "motivated" the publication through some proceedings.  In fact, several proceedings can appear even in  (or as) scientific journal issues, - which are not necessarily indicated as "special issues" by the publisher.  
Sometimes,  some ambiguity occurs on whether a publication pertains to a scientific report presented at a conference or is a more elaborate paper. 
In doubt, the publication was considered to be in the "proceedings" list. Note that  books written, translated,  or  proceedings edited  by the considered scientist are not counted here as relevant data. 

\begin{figure}
\includegraphics [height=10.8cm,width=10.8cm]{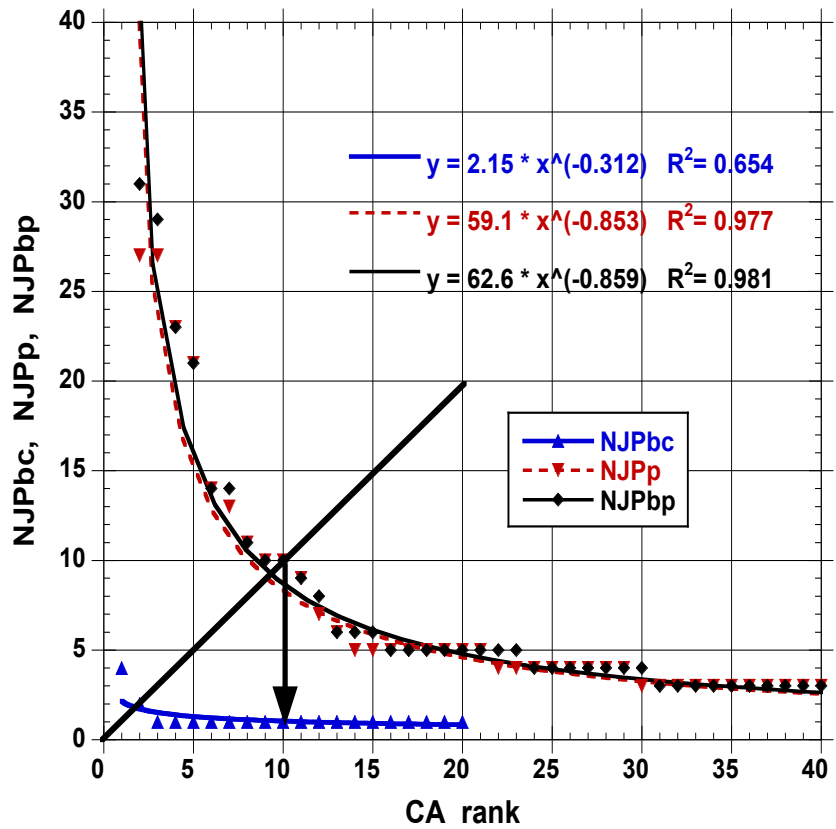}
\caption{   Number of Joint Publications (NJP) of   MA  with coauthors ranked by decreasing importance, in books and encyclopedia (NJPbc),  in proceedings 
 (NJPp),  as well as the "resulting sum"  (NJPbp),  in the vicinity of the coauthor core measure \cite{Sofia3a}, shown by the diagonal;  the corresponding $m_A$ value  (=10) is  indicated by an arrow;  each best fit is  given for each  $whole$ range  ($ r_M$,  NJPmfCA)  
  } 
\label{fig:Plot2maNJPbpbplinlin40}
\end{figure}

Thus, papers for encyclopedia and book chapters were first sorted   out, see Table 1, as  specific groups, (e) and (bc).  It turns out that  such contributions   are not in  great  numbers.   Moreover a test of Eq. (1) over such publications of e.g. HES, see Fig. \ref{fig:Plot9hesNJPpbpblinlin40}  and MA, see Fig. \ref{fig:Plot2maNJPbpbplinlin40}, indicates  that  the numerical effect of NJPbc is a marginal  contribution on NJPp, with respect to NJPbp, whether the whole range of $r$ or the central region of $J(r)$, near $m_A$ is used for  a power law fit.   Observe that  the  corresponding $\alpha$ exponent is close to, but above,  +1 in the HES case, and close to  +0.86  in the MA case.    Moreover the precision of the fit is rather  acceptable  in the case of the overall regime fits, i.e. $R^2 $ $\sim 0.873$  and $\sim$ 0.981, for the HES and MA case respectively, interestingly indicating a slight departure from a perfect power law in the HES case. However the fit is much improved, i.e.   $R^2$ $ \sim 0.
98$ for the HES case if  the central region only  ($r\le$ 30) is used (figure not shown).  This is shown for MA on Fig.   \ref{fig:Plot9maNJPpbpblog50}: $\alpha$ goes up to 0.88, with $R^2$ somewhat going down to $ \sim 0.96$. The variations can be considered to be weak, but in both cases $\alpha\neq$ +1.  One might easily understand that a size effect, the number of data points, is implied in such variations. The case of the other authors is not shown for lack of space, but if shown would not bring any different qualitative information.

Therefore,  it was decided to group all publications in two sets only, {\it in fine}  merging the (bc) and (e) into a (generalized) "proceedings" (p) list,  i.e. NJPbp $\rightarrow$ NJPp to be contrasted to  NJPj for the peer review journal (j) list. However, it was made sure that they are  all sound scientific papers, sometimes quoted in  works which are admitted to be regularly accepted as "papers".  Nevertheless,   e.g.,  it was found after completion of the analysis in Sect. \ref{sec:dataanalNJPp}   that  \cite{PhA375.DNA} which appears  at first to be a regular article in a well known journal, but (after reading it) turns out to be a conference paper, was misplaced.   Note that when attempting to remove ambiguities, it was considered that  one could  distinguish between papers published in either hard  or soft cover books. But journals are also soft cover\footnote{For example, in the case of MA, a JP, i.e., \cite{LOWT200IUPAP}, has been considered as a book chapter.}. Maybe other similar cases 
could be found. One must admit that one cannot read all papers to find out whether such a misplacement exists.  However, most of the analyzed data originates from deeply checked data bases. 
Note that  care was made   $not$  to count twice  identical scientific publications. For example, HES has many papers reproduced,  e.g. in a compendium or in another journal.  Moreover, an $Erratum$ has not been counted as a paper, but a  {\it "Reply to ..."} or a {\it "Comment on ..."} has been considered  a {\it bona fide} paper.   Overall, the error in each counting has been estimated to be at most  2.5\%.

Great care has been taken with the misprints of  CA names:  e.g.,    Buldryev and Giovanbattista,   are surely  Buldyrev  and Giovambattista, respectively.  
Great care has also been taken concerning polish, spanish, german, chinese and korean names.  First  (given) names and middle names, the latter sometimes missing, have been checked:  e..g., F.W. Starr and F. Starr are the same person; so is A. Petersen and A.M. Petersen.
Such a (tedious) manual check  has also allowed me to distinguish name homonyms,  like Ch. Laurent and Ph. Laurent.
HES also mentions  that one of his most often quoted   paper is attributed to some HFS!  All such  "defects" and   misprints  have been {\it a posteriori} corrected before manually counting the authors.  Though the paper grouping might be debatable, as well as the author selection, the lists can be considered as  safely reliable for further analysis within the present framework.

The statistical characteristics (mean, median,  standard deviation, skewness, kurtosis, ...)  of the various distributions $J(r)$ are not given, but they are available  upon request. 

Notations for the following text and Tables are as follows :
\begin{description}
\item[NJPmfCA :] number of joint publications with the most frequent (mf) coauthor;
\item[NJP1CA :] number of coauthors having written only 1 publication with  the LI  (whatever the number of coauthors on this publication); 
\item[TNCA :] total number of  coauthorships, whatever the CA frequency and number of publications; i.e.,  it is the integral of the histogram $J$  $vs$.  $r$;
\item[NDCA :] number of different coauthors.
\end{description}
   
\section{Analysis and Discussion of the  data set }\label{sec:data}

\begin{figure}
  \includegraphics [height=10.8cm,width=10.8cm]{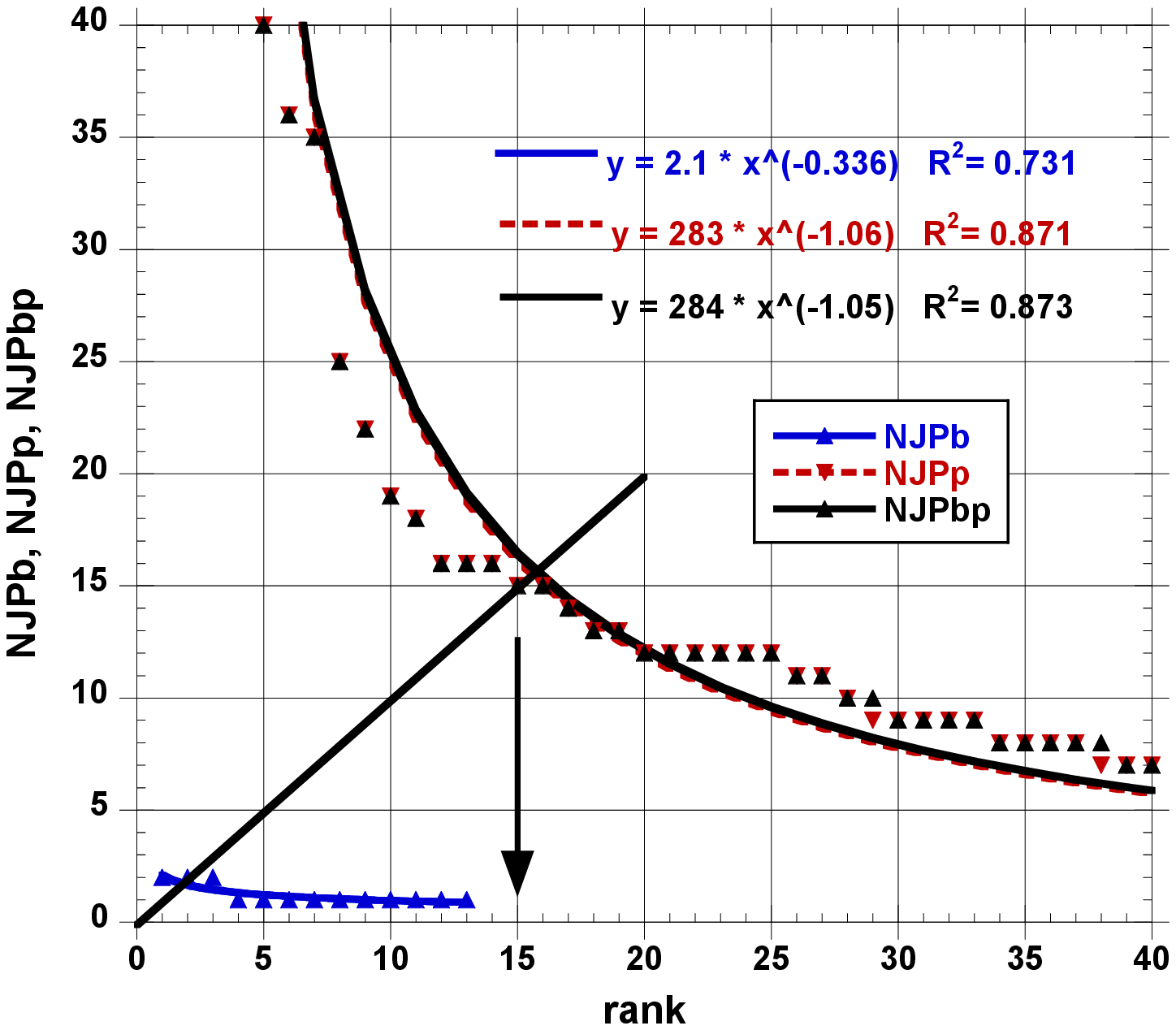}
\caption{   Number of Joint Publications (NJP) of   HES  with coauthors ranked by decreasing importance,  either  in books and encyclopedia (NJPbc) or in  proceedings   (NJPp),  and the "resulting sum"  combination  (NJPbp) in the vicinity of the coauthor core measure \cite{Sofia3a}, shown by the diagonal;  ; the corresponding $m_A$ value (= 15)  is indicated by an arrow;
 each best fit is  given    for the  $whole$ $r$ range  } 
\label{fig:Plot9hesNJPpbpblinlin40}
\end{figure}

HES publication list amounts to more than 1100 "papers", broken into subgroups, as  in Table 1. Its Curriculum Vitae \& Selected Publications, taken on  $polymer.bu.edu/hes/vitahes-messina.pdf $,  lists, among other things,    15 book chapters\footnote{ HES mentions only 14 book chapters, but I have considered that \cite{HESextrabookchap}, listed by HES as a journal publication, belongs to the book chapter category. As mentioned in the text, the sets (bc) and (e) are merged with the true proceedings one, but the category choice   in this case  has not much impact, as seen in Fig. \ref{fig:Plot9hesNJPpbpblinlin40}.}   and 5 encyclopedia articles,  619 articles, in the period 1966-1999 plus   more than 490 journal articles in the period 2000-up to  2012.   [Listed in rank order by citation count].  Note that his curriculum vitae (CV) mentions 136 Research Associates and ÒVisiting ScholarsÓ, at the end of 2012.   DS has more than 600 publications, several of them in German,  sometimes like comments. Several of  
his papers appear in  society journals or pedagogical ones and in not so usual peer review journals. In so doing, he has a larger (than many other scientists in the field) number of single author (sA) papers; see Table 1.  For HES and DS, in several cases,  sA or JP contributions stem from invited lectures published for anniversaries. Such papers are counted in (p). MA has  about 600 publications, mixing papers in peer review journals, proceedings (p),  contribution to encyclopedia (e), and book chapters (bc). In the DS and MA cases, some ambiguity occurs because several contributions pertain to papers derived from summer school lectures.  They were included in (p).  Two polish workers (JMK and AP) have a similar career age as HES, DS and MA. One  (JMK)  has a career partially in the USA. The 4 others are younger,  and essentially have a Poland based career.   Yet, DG started his career in high energy physics; he has a  paper in proceedings with 93 CA.  From a gender point of view,  all except KSW are male. 
With such a choice of authors,  the career data  indicates  different orders of magnitudes for the size of the publication list,   see Table 1.  Such a choice allows to cover  a large range of  coauthorship "types".

For comparison, the  characteristics of single author and joint publications of the 10 scientists so considered are given  in Table 1.  Note that in Table 1, the single author  (sA) papers are distinguished from the joint publications (JP), either in peer review journals (j) or in other categories (p, bc, e).  It is remarkable that the number of joint publications is much larger than the number of single author publications; the latter amounts to 5 to 10 \% only. The same order of magnitude, i.e. about  40\%, holds for the ratio  NJPp/NJPj for HES and MA.  However, this ratio is near 15\% for DS, AP, KSW, and JM. In contrast DG, RW and JMK have a much higher ratio ($\sim0.7$) than all others. Note, although this is a point outside the present discussion, that most of the NJPj or NsAPp  are $not$ often quoted, - as exemplified in any study based on the $h$-index!.  Since it takes some time before  a publication is quoted in some other work,  one should consider that some non negligible error bar exists on 
such a measurement of citations,  thus on the $h$-index, - in contrast to the present considerations on coauthors, who are "stacked to" the publication for ever. 

A rough estimate indicates that about 20\% of the JPp are $not$ quoted. An interesting Pareto number  order of magnitude!

The data emphasizing CAs is  summarized in Tables 2-4, grouping the papers into "proceedings" (p) and journal (j) types and some is displayed through  a few figures.   
In order to test Eq. (1), the overall $r$ range, i.e. $\in [1, r_M]$, has been used for  each fit. However  the $R^2$ value was not always found to be $\ge$ 0.9 (as 	already demonstrated in Fig. \ref{fig:Plot9hesNJPpbpblinlin40}). Therefore  a fit to a power law like Eq. (1) was often attempted in a restricted range, near the   region of interest, i.e. near $m_A$, such that $r$ $\in [1, r_{cf}]$ and  $J$ $\in [1, r_{cf}]$, with  $r_{cf} \sim 3m_A$, see each  figure caption for more information on the specific $r_{cf}$ value which is used.  This emphasizes the most frequent CAs, i.e. the  low $r$  regimes, but puts some limit on NJPmfCA, i.e. the very high $J$ cases. Two types of axis scales, i.e. lin-lin or log-log, are used, -- but the fits are  not always shown twice. The fit range is also indicated in  each figure caption.   For better vizualisation, only the low $r$ range ($<3m_A$), and by symmetry, the range $J\le3m_A$ data is shown. Examined authors have been grouped  on   illustrating plots   such 
that the data can be read as  easily as possible, even though it it unavoidable that  they   do   often overlap.
 
\subsection{NJP Numerical Analysis}  \label{sec:dataanalNJP}

\begin{figure}
\centering   \includegraphics  [height=10.8cm,width=10.8cm] {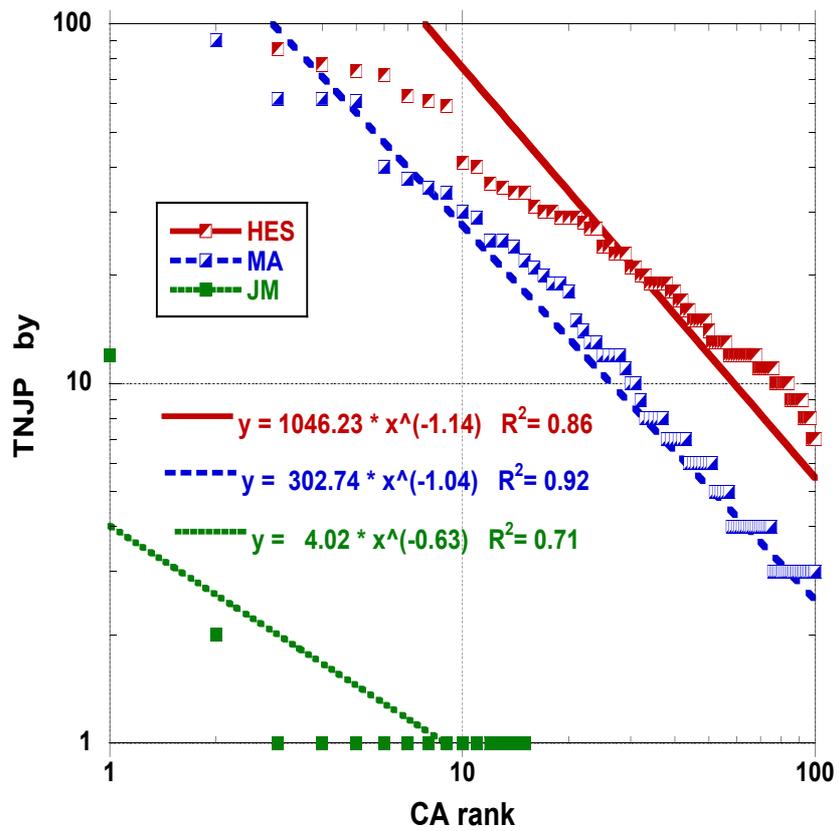}
\caption{ log-log scale display of the Total Number of Joint Publications (NJP) by HES, MA, and JM, with coauthors  (CA) ranked by decreasing importance  
in the vicinity of the  so called Ausloos coauthor core measure  \cite{Sofia3a};    
 best fits are given   for the  $whole$  $(r,J)$ range  }
\label{fig:MAHESb}
\end{figure}

\begin{figure}
\centering  \includegraphics  [height=10.8cm,width=10.8cm] {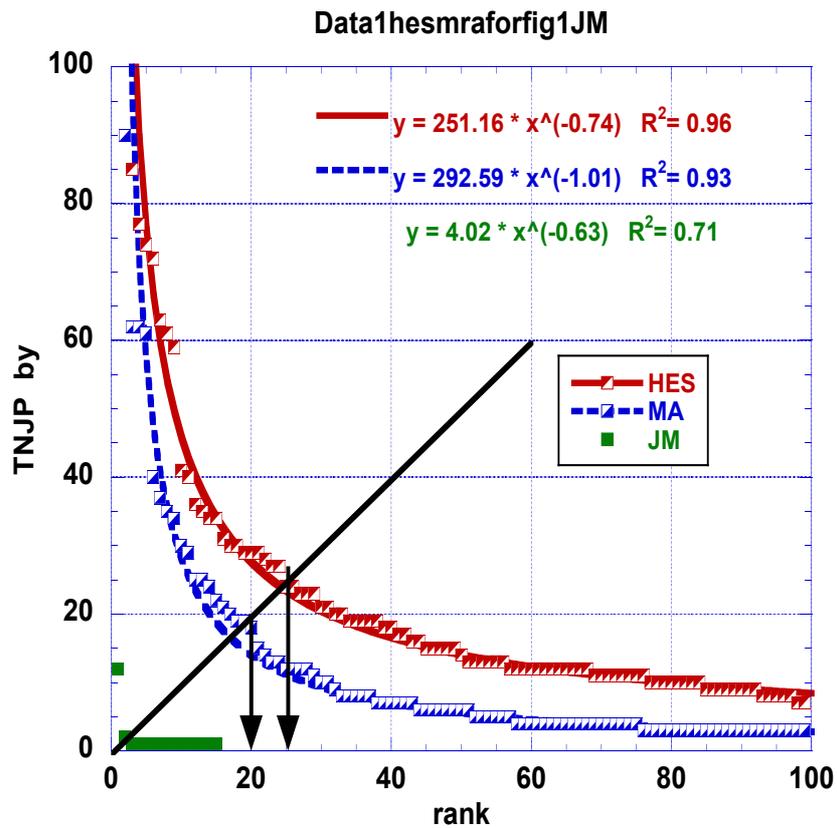}
\caption{    Total Number of Joint Publications (TNJP) by HES, MA, and JM, with coauthors  (CA) ranked by decreasing importance  in the vicinity of the  so called Ausloos coauthor core measure  \cite{Sofia3a}, shown by the diagonal;  TNJP $m_A$ values for  HES and MA  are indicated by arrows;  
best fits are given for the $central$ region only, i.e. both $r$ and $J$   $\le$ 100   }  
\label{fig:MAHESa}
\end{figure}

Not all data and fits can be displayed, but the relevant results are summarized in  Tables, 2-4. 
A few  (3) cases of  the Total Number of Joint Publications (TNJP)  data  ranked according to the CA importance are displayed and numerically fitted on  Figs.  \ref{fig:MAHESb} -  \ref{fig:MAHESa}. 
Observe that   $\alpha$ (TNJP) is close to +1 for HES and MA (Fig. \ref{fig:MAHESb}).  The $R^2\sim 0.9$ and $\alpha$ values are  thus reliable whatever the range fit, i.e. on the  whole range or near $m_A$, surely for MA, but note that $\alpha$  for HES falls from 1.14  down to 0.74 if the range is reduced.   Observe also  that   the  overall $\alpha$ (TNJP)  falls below 0.8 for DS and JMK, and near 0.7 for KSW and RW.  
 
The $m_A$ core value is found, - e.g. as indicated for HES and MA by an arrow in Fig. \ref {fig:MAHESa}  and for DS, JMK and AP in Figs.   \ref{fig:Plot22TNJPDSJMKAPlolocol}, \ref{fig:Plot30155},   \ref{fig:Plot21DSJMAP30x30lilicol}. The  TNJP  $m_A$ for all authors are  also given in Table 2-4. The values range from 2 to 26. 
   
Observe at one extreme, i.e.  JM,  who has a small number of  frequent CAs,  together with a small number of data points, the  CAs with $r\ge m_A$, - all of the 1CA type,  much constraint the "fit",  - factually  decreasing  $\alpha$  to a lower value, i.e. 0.63 than the NJPj $\alpha$. This is somewhat similar to the case of DS for which the $m_A$ value does  NOT increase from its NJPj  to its TNJP value. Finally  the case of DG is  also  of interest  from the TNJP point of view, since it has the lowest $\alpha$, dragged down by  the NJPp behavior, as hinted here above. 

Such an observation  further indicates  the interest of investigating the effect of the number of CAs on specific types of publications, like the "proceedings", as done in the next section.

\subsection{NJPp Numerical Analysis}  \label{sec:dataanalNJPp}
  
\begin{figure}
 \centering  \includegraphics  [height=10.8cm,width=10.8cm]{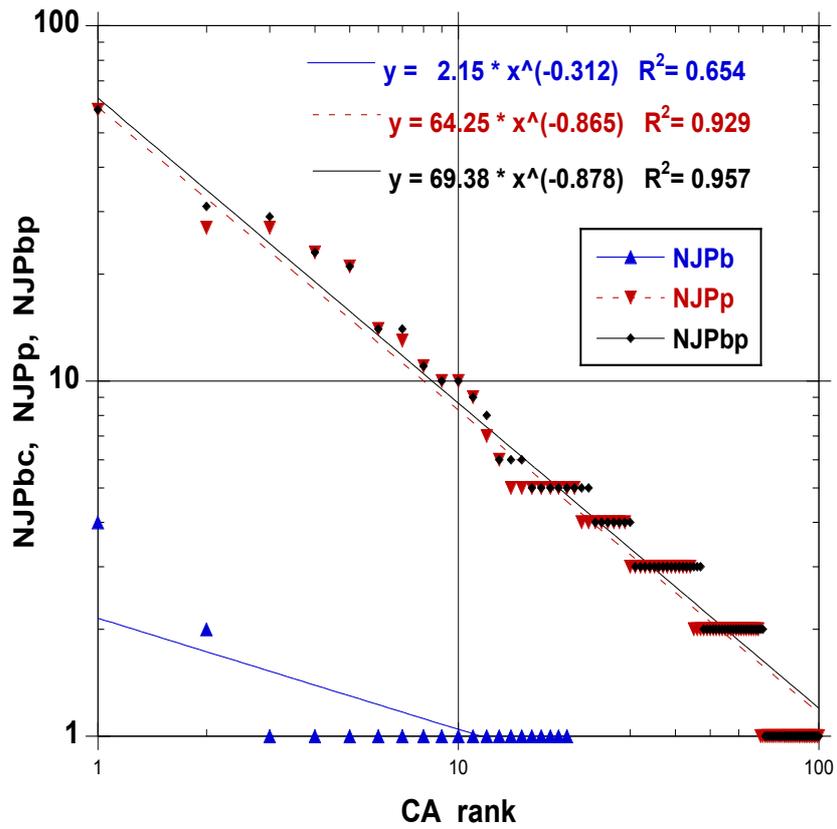}
\caption{  Log-log display of the Number of Joint Publications (NJP) of   MA  with coauthors ranked by decreasing importance, in  both books and encyclopedia (NJPbc) and in  proceedings (p)  (NJPp), and their  "resulting sum"  combination (NJPbp);
 each best fit is  made on the data $central$ $r$ range ($r \le40$, $J \le40$)}  
  \label{fig:Plot9maNJPpbpblog50}\end{figure}

\begin{figure}
 \centering  \includegraphics  [height=10.8cm,width=10.8cm]{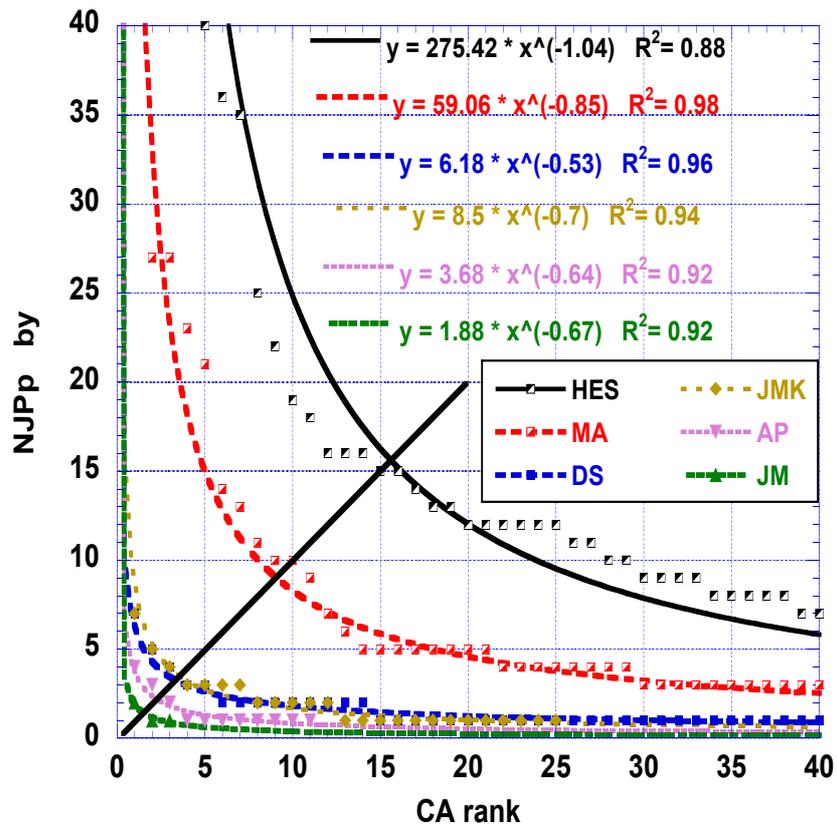} 
\caption{ Display of the Number of Joint Publications  in "proceedings"    (NJPp) by   6 LIs , i.e. HES, DS, MA, JMK,  AP,   and JM with coauthors ranked by decreasing importance, near $m_A$; its value, being suggested by the diagonal, is found in Tables 2-4;  each best fit is  given    for   the $whole$ $r$ and $J$ range   ($ r_M$,  NJPmfCA) } 
  \label{fig:Plot2NJP_6_40x40lili}\end{figure}

\begin{figure}
 \centering  \includegraphics  [height=10.8cm,width=10.8cm] {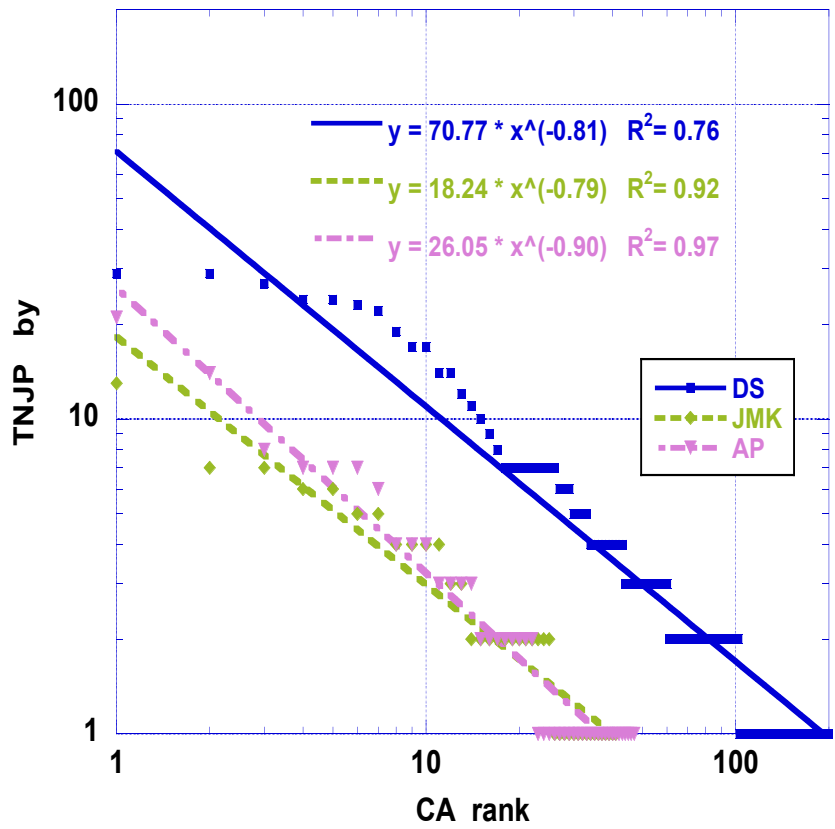} 
\caption{ log-log scale display  of the Total Number of Joint Publications (TNJP) by DS, JMK, and AP, with coauthors  (CA) ranked by decreasing importance   
 best fits  to a power law are given  for the  $whole$  $(r,J)$ range  }
  \label{fig:Plot22TNJPDSJMKAPlolocol}
   \end{figure}

\begin{figure}  
 \includegraphics  [height=10.8cm,width=10.8cm] {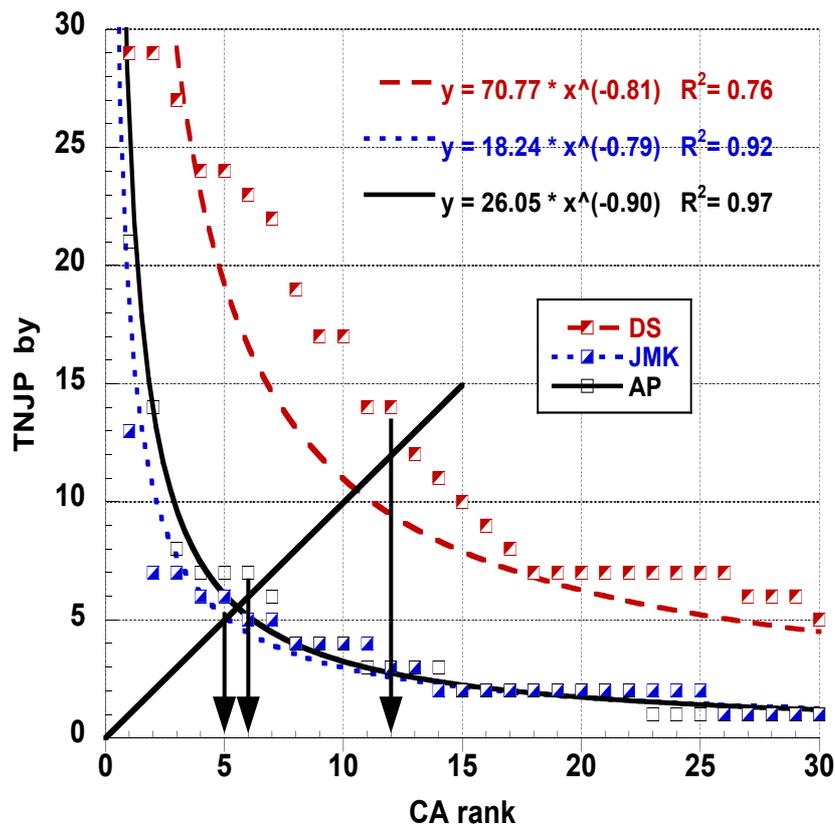} 
 \caption{Total Number of Joint Publications (TNJP) by DS, JMK, and AP, with coauthors  (CA) ranked by decreasing importance;  each best fit  to a power law is given 
 for the    whole $(r,J)$ range; the so called Ausloos coauthor core measure  \cite{Sofia3a} is indicated by an arrow for each LI }   \label{fig:Plot30155}
 \end{figure}

\begin{figure}  
 \includegraphics  [height=10.8cm,width=10.8cm] {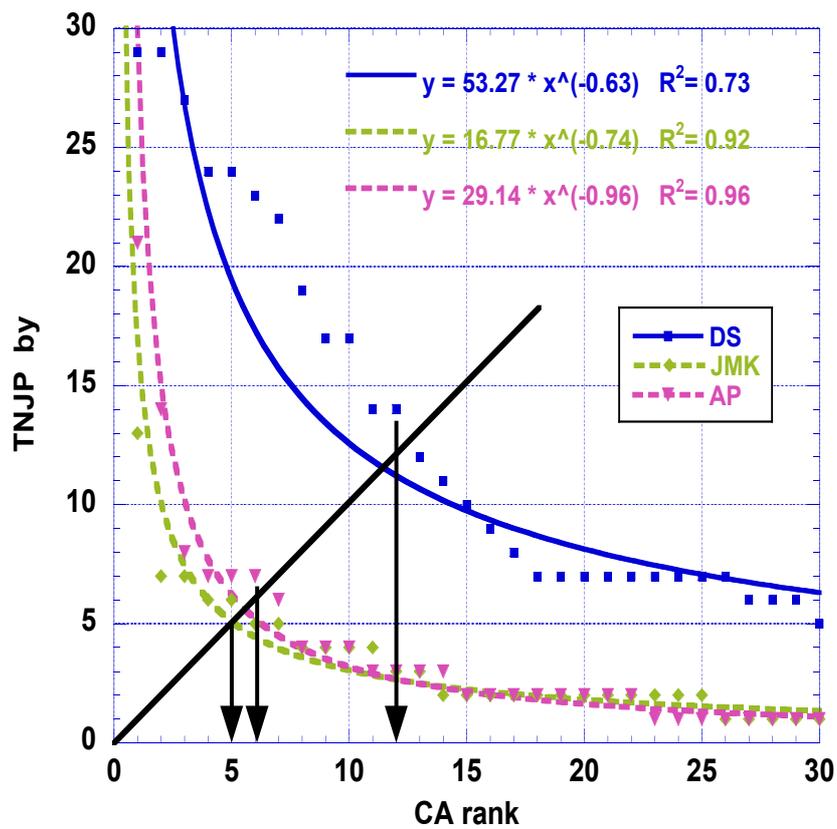}  
\caption{Total Number of Joint Publications (TNJP) by DS, JMK, and AP, with coauthors  (CA) ranked by decreasing importance;  each best fit  to a power law is given 
 for the    [30,30]  $(r,J)$ range, i.e. near the $m_A$ index,  i.e., the so called Ausloos coauthor core measure  \cite{Sofia3a}, indicated by an arrow for each LI }  \label{fig:Plot21DSJMAP30x30lilicol}  
  \end{figure}

Several plots of the NJPp behavior have already been on  Figs. \ref{fig:Plot9hesNJPpbpblinlin40}
 -  \ref{fig:Plot9maNJPpbpblog50}. Others are presented in Figs.  \ref{fig:Plot2NJP_6_40x40lili} -   \ref{fig:Plot21DSJMAP30x30lilicol}.
In Tables 2-4 one can find  the set of relevant numerical results.  
Interestingly,  the NJPp power law fits are much better  ($R^2 \sim$ $0.92$) for  almost all authors (except HES and DG!) than for  the TNJP; see  Figs. \ref{fig:Plot2NJP_6_40x40lili} -   \ref{fig:Plot21DSJMAP30x30lilicol}.   
   
Let  it be observed whether the hyperbolic law, Eq. (1),  is obeyed or not, first,  in the HES case,  Fig. \ref{fig:Plot9hesNJPpbpblinlin40},  for NJPp. 
The fits for  NJPp are quite fine ($R^2\ge 0.87$), with $\alpha\simeq 1.05$, whatever the range considered.  
The $m_A$ value is well defined: $m_A=$15 for   NJPp.  Moreover,  the $\alpha$ value can be large, as for KSW, i.e. $\simeq 1.255$, and be low, as for DG, i.e. $\simeq 0.239$. 
  
A  test has been made on several cases concerning the fit range.  An example of the central range fit is shown in Fig.  \ref{fig:Plot9maNJPpbpblog50}, for MA, Fig.  \ref{fig:MAHESa},  for HES and MA, and Fig. \ref{fig:Plot21DSJMAP30x30lilicol}
for  DS, JMK and AP.    The numerical difference  on $\alpha$, between fits with different ranges, is not large
The NJPp  MA value of $m_A$ is found to be equal to 10.  
   
However some deviation from a nice hyperbolic  fit  is apparent   in the vicinity of $m_A$, - indicated by an arrow. Nevertheless the $m_A$ values are always well defined.
  
In all  cases, it can be observed that the   $m_A$ values  are rather small : $m_A$ = 15 and 10 for HES and MA respectively, but fall below 4 for all others.   In contrast DS,  has a very low NJPp $m_A$ value   (= 3),  and  a remarkably low $\alpha$ value ($\sim 0.5$).   In fact,   similarly low  $m_A$ and $\alpha$  values are also found for the polish workers. All  NJPp $m_A$ values fall much below the   NJP  $m_{A} $ coauthor core value, i.e. $m_A$ = 25 and 20 for HES and MA respectively, as  recalled  in  Table 2.  
    
\subsection{  Discussion  }\label{sec:dataset}
  
The anomalous behaviors, i.e. large deviations from the expected Eq. (1) law,  maybe on one hand attributed to the data itself and to the approximation of a simple scaling law. On the other hand, it might reflect some deeper information on the LI role, leadership, management,  funding, research topics, ..., and team organization, history, constraints  \cite{HESPRE2010career,HESPNAS2012career,HESPNAS2011career}.
   
First, indeed, one can expect to observe, as in many Zipf plots, some irregularity at low $r$ and at large $r$. A so called "queen effect" \cite{Sofia3a} indicated by a sort of horizontal curbing in the data,   see Fig. \ref{fig:MAHESb} or  Fig. \ref{fig:Plot22TNJPDSJMKAPlolocol},
could be  represented by a  hyperbolic  Bradford-Zipf-Mandelbrot-like law,   
 \begin{equation}\label{ZMlikeCr}
  J =\frac{J^{*}}{(\nu+r)^\zeta},
 \end{equation}
expectedly with $\zeta\simeq+1$ \cite{FAIRTHORNE}.  In this low $r$ range, it is known that a so called "king effect"  \cite{EPJB2.98.525stretchedexp_citysizesFR}, i.e. a sharp upturn at low $r$ values  exists, if such a low ranked element has much preponderance.
In the case of HES, one also encounters a king and a queen effect, in several cases. The king effect is due to Havlin, Buldyrev,   Amaral,  Gopikrishnan,  and Plerou,  for $r= $ 1 to 5, those CA with NJPbp $\ge40$.  The CA order differs from that in  NJPj and TNJP (not shown). Similarly for MA, the king effect is Cloots who has markedly a NJPbp $>>40$, see Table 2.  
For MA, the CA rank order also differs from that in  NJPj and TNJP (not shown).
   
Some upturn seen in the case of HES, near $r >$  10  can be  attributed to the fact that HES has been very active in many different fields for which the CAs are somewhat  similar but with different emphasis, the more so due to the timing of his research. Such  cases are much concerned with medical topics. It is understandable that a team effect, with kings and queens , are to be expected in such domains. Such domain effects are  necessarily masked into the overall plot,  but allow  some distinction between papers in "proceedings" and those in peer review journals.
 
Finally,    the  fact that the  "proceedings" (and also the journal) $m_{A} $ values fall much below the   overall $m_{A} $ coauthor core value could be  interpreted   as only being due to  the size of the number of CAs. However, very similar kings and queens occur in different publication types. The above finding may suggest that $m_A$ is topics dependent. A normalization, leading to a frequency plot, might be  useful  work in order to improve $m_A$ as a standard measure.  However another interpretation might also pertain to the type of publication itself. In fact, it can be observed that the approximatively same value is found for the ratio NJPp/NJPj , for  the TNCA for both HES and MA, i.e. $\sim 0.47$. However this ratio, either for the NJP1CA or the  NDCA, are very different for HES and MA. The values for the HES case are about half those for the MA case. Since NJPp is (very) roughly the same for HES and MA, see Table 2, this necessarily indicates a quite different emphasis and success in the 
publication of scientific papers with CAs,  in peer review journals by HES and MA.
   
From its NJPp $m_A$ value,  as well as its low $\alpha$ value, one may conclude that  DS is  surely a special (top) scientist, in this respect.
 
\section{Time regimes }
\label{sec: Appendix}
 
\begin{figure}
  \includegraphics  [height=10.8cm,width=10.8cm] {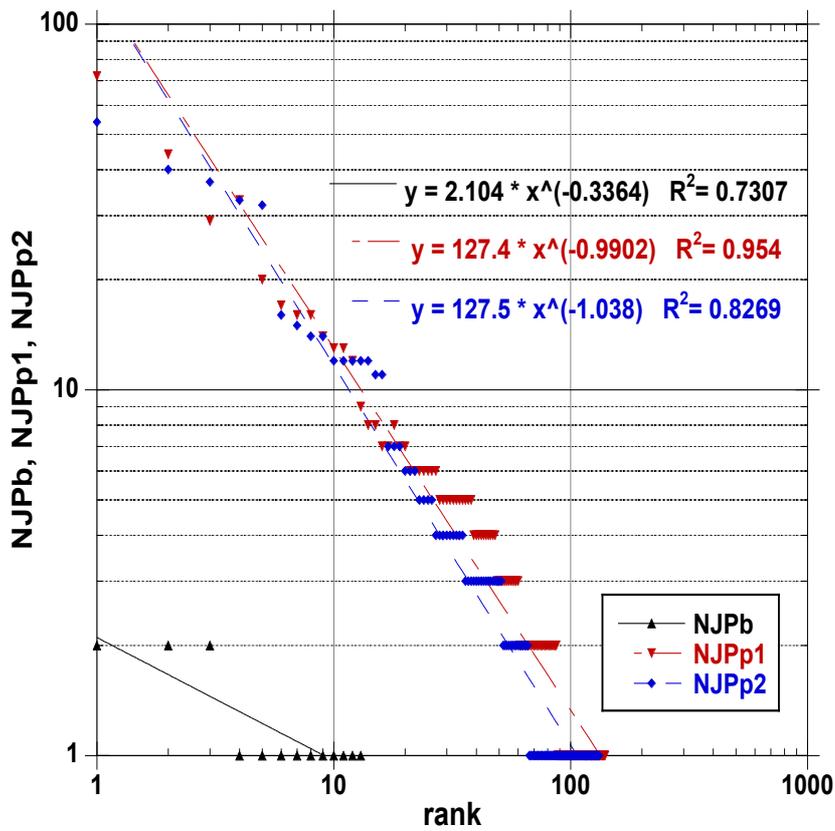} 
\caption{  Log-log display of the Number ofJoint Publications (NJP) of  HES  with coauthors ranked by decreasing importance, in books and encyclopedia (NJPb),  and  in "proceedings" (see text for definition) (NJPp), during different time spans (p1  and p2), see Table 5; each  
 best fit is given    for the  whole ranges   }
\label{fig:Plot1hesNJPp}
\end{figure}

\begin{figure}
 \includegraphics  [height=10.8cm,width=10.8cm] {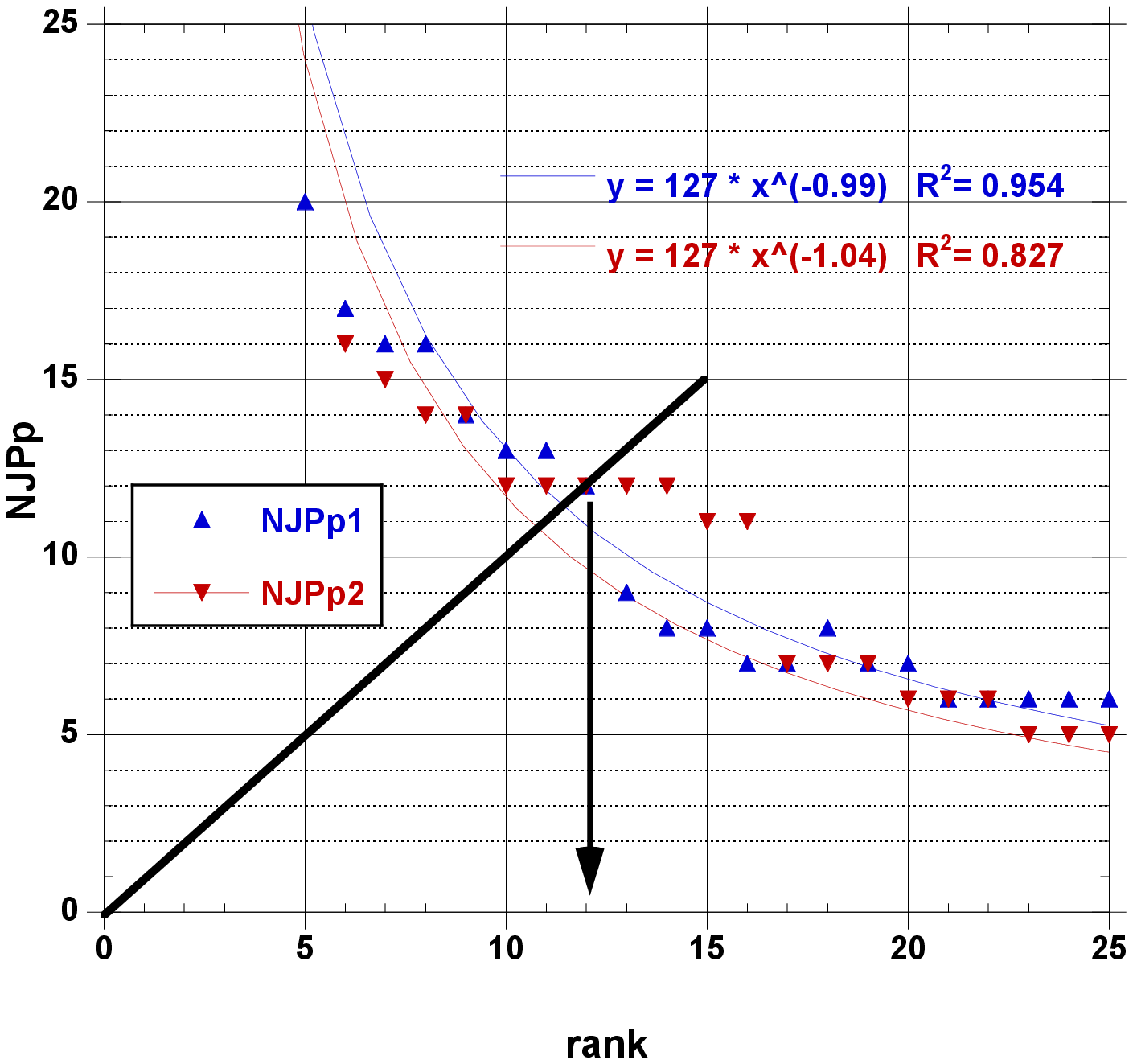}
   \caption{  Number of Joint Publications (NJP) of  HES  with coauthors ranked by decreasing importance,  in "proceedings" (see text for definition) (NJPp), during different time spans (p1 and p2), see Table 5, in the vicinity of the coauthor core measure \cite{Sofia3a}, shown by the diagonal;  $m_A$ value  (=12) is  indicated by an arrow;
 each best fit is  given    for the  wholeranges  } 
\label{fig:Plot1hesNJPplinlin25a}
\end{figure}

\begin{figure}
\includegraphics  [height=10.8cm,width=10.8cm]{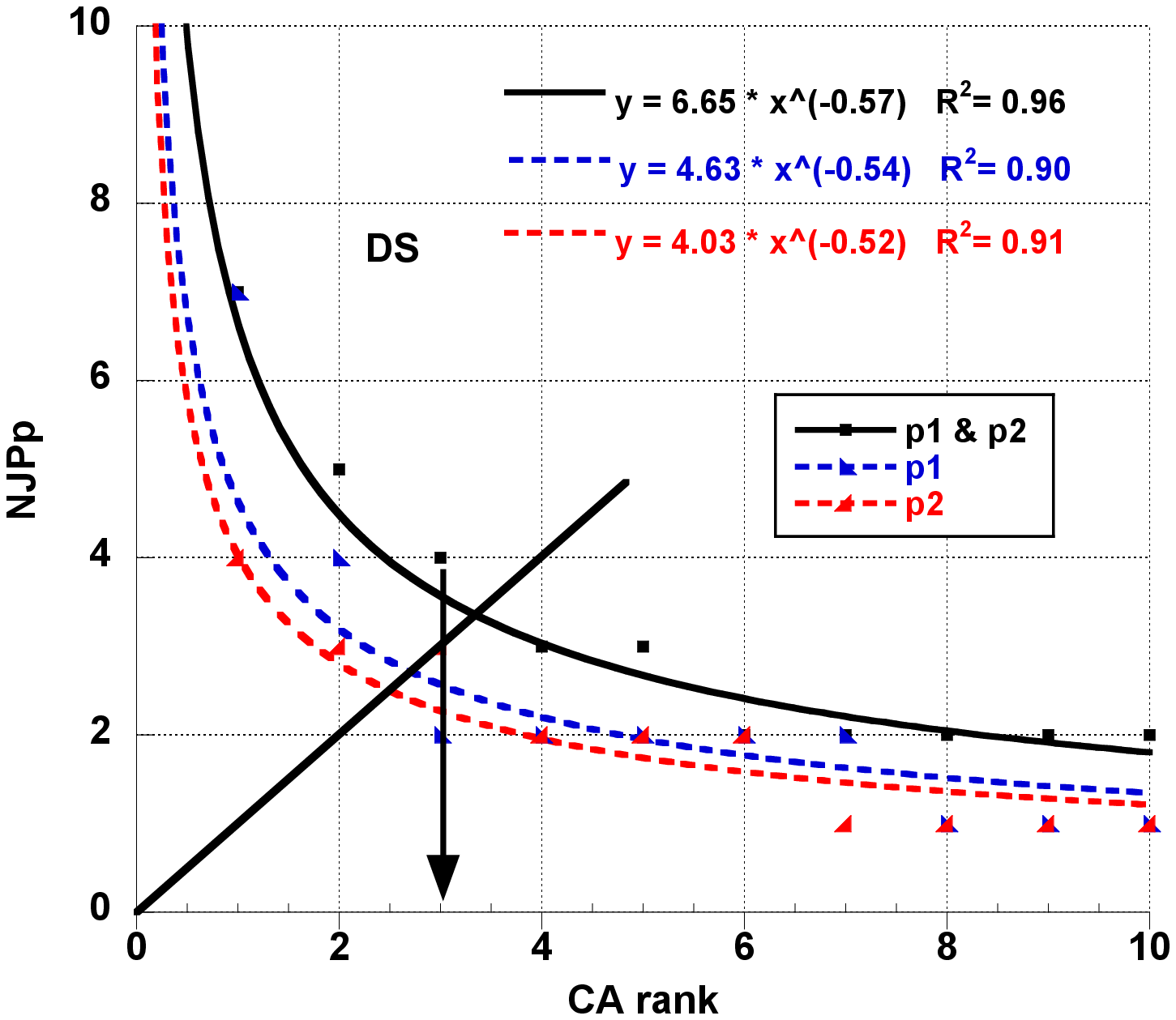}
 \caption{  Number of Joint Publications (NJP) of  DS  with coauthors ranked by decreasing importance,  in "proceedings" (see text for definition) (NJPp), during different time spans (p1 and p2), see Table 5, in the vicinity of the coauthor core measure \cite{Sofia3a}, shown by the diagonal;  $m_A$ value  (=3) is  indicated by an arrow;
 each best fit is  given    for the  whole  ranges  }  
\label{fig:Plot6DSp1p2plili10}
\end{figure}

Since HES decomposes his publication list in his CV into two regimes [1966-1999] and [2000- ...], it is interesting to find out whether there is some sort of time dependence of the above findings. On one hand, the scientific subjects have shifted; on the other  hand, the coauthors might have also changed. Moreover the scientific output of research teams might have changed due to the use of many electronic means and other funding constraints. Thus,  in
Table 3,  the HES and DS JP characteristics are reported for both "time regimes", while Fig. \ref{fig:Plot1hesNJPp} and Fig. \ref{fig:Plot1hesNJPplinlin25a} illustrate the data in the case of HES and Fig. \ref{fig:Plot6DSp1p2plili10} in the case of DS.     Interestingly, the  NJP, NPmfCA, NDCA, NCA, $\alpha$, $R^2$, and $m_A$ values do not seem to depend on the time regime, thus on the emphasis on one or another topics, nor on the CA.  Surprise, surprise.

It is seen that NsA is different in the different regimes: this can be understood  as follows: at the beginning of a career, it is more probable that either  a (young) scientist belongs to a big group and thus at once gets many coauthors, or he/she searches for recognition, thus in publishing sA papers, - when or if allowed to do so. 
 
\section{Conclusions}
\label{sec:conclusions}

Scientific production and other  facts,  like citations and (joint) publications,  are used nowadays for quantifying  scientific achievements 
\cite{HESPRE2010career,HESPNAS2012career,HESPNAS2011career,hindex,hirsch10,PriceModelSgrowth2004} of a scientist or a team.  Yet everyone admits that  statistical methods based on mere arithmetic counts are only partially adequate  because any quantitative bias omits relevant qualitative features  \cite{OrmerodOmega28} and  the counting is  sensitive to data size, while  interactions and contextual variations are somewhat hidden\footnote{  Let  the  work on   the critical mass and the dependency of research quality on group size by    Kenna  and  Berche  \cite{1006.0928criticalmasskenna} be mentioned.}.  A flurry of papers have been produced  "to improve" these "measures", like (not an exhaustive list!)   \cite{Radiology255.10.342 bibliom,Schreiber2010b,Schreiber2012JoI6_17v,0912.3573 zhang,JASIST59.08.830BornmannMutzDaniel}. 
  
It is somewhat of common knowledge that proceedings papers contain more coauthors than peer review journals. This is exemplified here in the ratio NDCA/TNCA, either for NJPj or NJPp; see Tables 2-4. Officially coauthorship should  imply personal responsibility for the content of  a paper.
Yet,   undeserved authorship, assigning authorship to persons because of their authority or prestige, or as courtesy,  seems much allowed in proceedings papers, resulting from conference presentations. Along with increasing number of coauthors, the percentage of undeserved (false) authorship  has been shown to  increase  \cite{8,9}. Conversely,  some sense of "obligation", fear of offending someone, need for justifying some collaboration, pressure from another coauthor, or explicit demand, gaining favor or reciprocity \cite{Martin1304.0473v1authorcitation}, are common facts.  

Ausloos coauthor core definition and subsequent measure \cite{Sofia3a} tackle such considerations in a constructive way, through the relationship between the number ($J$) of (joint) publications with coauthors ranked according to rank ($r$) importance. The approach presents a great difference  with respect to the Hirsch index. The former emphasizes the role of persons.  The latter test the popularity of a paper. 

A test of the findings  in \cite{Sofia3a}, i.e.,    $ J \propto 1/r^{\alpha}$, with $\alpha\simeq 1$,  has been made   here above and discussed considering two prolific highly respected authors, often quoted, and  several others working in the field of statistical physics.  
The role of coworkers, in fact coauthors, has been examined supposing that publications in peer review journals  and in so called proceedings  might have some different influence on the core of coauthors of a LI.  This is proved to be  so  for the here above examined cases.   It is observed that  the $R^2$ fit values are usually quite good, in particular considering that a mere power law fit is attempted.  The $m_{A} $ values of the so called coauthor subcores fall below the overall $m_{A} $ coauthor core value, and  differ  whether publications  are in peer review journals  or in so called proceedings. Practically,  the $\alpha$ values seem to be size and  publication type dependent as well.  Inn conclusion, contrary to the universal value somewhat thought of    in \cite{Sofia3a},  it seems that $\alpha$ and $m_A$ are scientist dependent. Thus practical considerations on a LI and  research team behavior can  be observed through the TNJP exponent  $\alpha$  and the coauthor core value $m_{A} $.   

Whence, much can still be done after the above, in particular, checking  (i) the deviations from a regular hyperbolic law; (ii)  the  the king and queen effects; (iii) the effect of NDCA   long tail; (iv) the influence of  NJP1CA; (v) the effect of the number of CAs per paper\cite{[35],[36]}; (vi)  globalization of measures in considering the role of the main author, and in the ranking of his team mates.  Some modelization is left for further work.  One could imagine extensions of Egghes success-breeds-success model \cite{egg1}, leading to not necessarily decreasing distributions.  

 Technically, one could thus measure the relevant strength of a research group centered on some leader, e.g. through the various ratios discussed in the main text. Nevertheless, the invisible college  \cite{HK94,Zuccala06invisblcoll}    becomes visible and easily quantified,  including hubs in so doing.  One may conjecture that one can introduce selection and rewarding policies in the  funding of a team (or LI)  through  Ausloos coauthor core measure   $m_A$    \cite{Sofia3a}. To indicate   applications toward practical consequences, e.g. on team (or LI) funding or  criteria for career evolution,  are outside the scope  of this paper.

\bigskip

{\bf Acknowledgements} Thanks to M. Ausloos for private communications on \cite{Sofia3a}, comments prior to manuscript submission and making available his  publication list broken according to the present investigation.

\newpage

     \begin{table}\label{incomexpensfits}
      \begin{tabular}{|c||c|c|c|c|c|c|c|c|c| c|    }
  \hline
&\multicolumn{9 }{|c|}{data summary}    \\
\hline     &NP&NJPj&NsAPj&NJPp&NsAPp&NJPbc&NsAPbc&NJPe&NsAPe    \\
\hline
\hline    HES&1148&791&15&288&34&9&6&2&3    \\
 \hline
 \hline     DS & 612&374&115&46&77&-&13&-& 4   \\
 \hline
 \hline  MA &599&359&29&148&31&15&5&1&1   \\ 
\hline 
 \hline     RW&98&32&7&21&5&11&1&0&0    \\
 \hline
 \hline     JMK&60&28&8&19&5&-&-&-&-     \\
 \hline
 \hline
 \hline     AP &111&79&22&9&1&-&-&-&-    \\
 \hline
  \hline    DG&45&12&14&9 & 10&0&0&0&0    \\
 \hline
  \hline    KSW&43&27&7&4 &5&0&0&0&0    \\
 \hline
 \hline    JM & 27 & 15 & 8 & 2 & 1 & 1 & 1 & 0 &0     \\ 
\hline \end{tabular} 

\caption{ NP: total number (N)  of publications (P);  
NJPj: number of joint publications (JP)  in  journals (j); NJPp: number of joint publications in proceedings (p);  NJPbc: number of  joint publications in book chapters (bc); NJPe: number of joint publications in encyclopedia  (e);  sA: single author; data updated  in Dec.  2012 }
\end{table}

\begin{table} 
 \begin{tabular}{|c|c|c|c|| c|c|c|| c|c|c|   }
   \hline
&\multicolumn{3}{|c||}{HES}&\multicolumn{3}{|c||}{ DS}&\multicolumn{3}{|c|}{ MA} \\
\hline
\hline &  NJPj& NJPp& TNJP &   NJPj&NJPp & TNJP  &  NJPj &NJPp& TNJP   \\
\hline 
\hline   oldest P &1965& 1969 & 1965   &   1968 &1973&1968&   1971 &1971 &1971  \\
  latest P& 2012& 2012& 2012 &   2012  &2011  &2012 &   2012 &2011&2012\\
\hline NJPmfCA	 &195&104&299	&27&7&29&	97&58 &155 \\
NJP1CA &275&102&269&172&32&178&152&103&172   \\
TNCA 	&2639&1250&3889	&691&72&763&1055&502&1557 \\
NDCA	&568&242&592&268&46&280 &	273&	168&319\\ \hline
\hline
\hline  $\alpha$    &  0.999& 1.045& 1.135 &    0.688&0.535&0.796 & 1.03&0.860 &1.04    \\
  $R^2$  & 0.914& 0.873& 0.86  &  0.752&0.965&0.722 &  0.91&0.981 & 0.92    \\
 $m_{A} $&20 &15&26&	12&3&12&15 &10 &20  \\ \hline
\end{tabular}
\caption{Summary of  data  characteristics  for the number of  joint publications (NJP) of HES, DS and  MA according to notations given in  text, Sect. \ref{sec:dataset}, i.e. distinguishing journals (j), "proceedings" (p) and the  total (T) sum (TNJP). The CA law    core characteristics, $\alpha$ and $m_A$ \cite{Sofia3a}, are also given from fits discussed in the text and figure captions }\label{Tablestat1}
 \end{table}  
 
 \begin{table} 
 \begin{tabular}{|c|c|c|c|| c|c|c|| c|c|c|   }
   \hline
&\multicolumn{3}{|c||}{ RW}&\multicolumn{3}{|c||}{ JMK}&\multicolumn{3}{|c|}{ AP} \\
\hline
\hline &  NJPj& NJPp& TNJP &   NJPj&NJPp & TNJP  &  NJPj &NJPp& TNJP    \\
\hline 
\hline   oldest P &1997& 1995 & 1995   &   1969 &1971 &1969&   1969 & 1980 & 1969   \\
  latest P& 2012& 2010& 2012 &   1999 &1999&1999 &   2012 &2010&2012\\
\hline NJPmfCA	 &5&10&15	&6&7&13 &17& 4&21 \\
NJP1CA &13&11&24&24&13&16& 25 &8&25   \\
TNCA 	&62&62&124&60&51&111& 118&17&135 \\
NDCA	&34&23&46&	35&25&41 	&45&11&47\\  \hline
\hline
\hline  $\alpha$    &  0.561& 0.767 &0.743&    0.618&0.702&0.787 &0.89 &0.64& 0.94    \\
  $R^2$   & 0.867& 0.901& 0.918 &  0.887&0.943&0.915 &  0.945 &0.92& 0.96    \\
 $m_{A} $ &4 &4&6	&4&3&5&5&2 &6 \\ \hline
\end{tabular}
\caption{ Summary of  data  characteristics  for joint publications of RW, JMK and AP according to notations given in  text, Sect. \ref{sec:dataset}, i.e. distinguishing journals (j), "proceedings" (p) and the  total (T) sum (TNJP). The CA law    core characteristics, $\alpha$ and $m_A$ \cite{Sofia3a}, are also given from fits discussed in the text and figure captions }\label{Tablestat2}
 \end{table}  
 
 \begin{table} 
 \begin{tabular}{|c|c|c|c|| c|c|c|| c|c|c|   }
   \hline
&\multicolumn{3}{|c||}{DG}&\multicolumn{3}{|c||}{KSW}&\multicolumn{3}{|c|}{JM} \\
\hline
\hline &  NJPj& NJPp& TNJP &   NJPj&NJPp & TNJP  &  NJPj &NJPp& TNJP    \\
\hline 
\hline   oldest P &2000&2000&2000  &   1996 &1999 &1999&   2001 & 2006 & 2001   \\
  latest P& 2012& 2000& 2012 &   2012 &2006 &2012  &   2012 &2006&2012\\
\hline NJPmfCA	 &5&5&10&	6&3 &9 &	10& 2	&12 \\
NJP1CA &4&3&7&10&0&10& 10 &0&10   \\
TNCA 	&14&104&118&44&5&49& 23	& 2 & 27 \\
NDCA	&7&99&104& 21&3&21&	12&	1&14\\ \hline
\hline
\hline  $\alpha$   &  0.755& 0.239& 0.547 &    0.594&1.255&0.715 &0.67 &0.67& 0.63    \\
  $R^2$  & 0.933& 0.321& 0.615 &  0.865&0.954 & 0.961 & 0.76 &0.92& 0.71    \\
 $m_{A} $&2 &2&2 & 3 &1&3&2 &1 &2  \\ \hline
\end{tabular}
\caption{ Summary of  data  characteristics  for joint publications of  DG, KSW and JM according to notations given in  text, Sect. \ref{sec:dataset}, i.e. distinguishing journals (j), "proceedings" (p) and the  total (T) sum (TNJP). The CA law    core characteristics, $\alpha$ and $m_A$ \cite{Sofia3a}, are also given from fits discussed in the text and figure captions }\label{Tablestat3}
 \end{table}  

 \begin{table} 
      \begin{tabular}{|c|c|c|c|| c|c|c|    }
  \hline
 &  \multicolumn{3}{|c||}{HES   }&   \multicolumn{3}{|c|}{DS   }  \\ \hline 
    \hline
 &     \multicolumn{3}{|c||}{time span regimes  }& \multicolumn{3}{|c|}{time span regimes  }   \\ 
\hline & [1966-1999]&[2000-2012]&[1966-2012]& [1966-1999]&[2000-2012]&[1966-2012] \\\hline 
 \hline& NJPp1 &  NJPp2 & NJPp & NJPp1 &  NJPp2 & NJPp  \\
\hline NJP&  150 &138& 288&   41&31&72 \\
\hline NJPmfCA&72&54& 104& 7&4&7 \\ 
\hline NJP1CA&53&66& 102 & 20&15&32\\ 
\hline NDCA  & 140&132&242 &    27&21&46\\ 
\hline NsA  &28  &8&36 & 46&32&78 \\  \hline
\hline $\alpha $& 0.99 &1.04 &1.045 &  0.54&0.52&0.565  \\ 
\hline $R^2 $ & 0.954&0.827 &0.875&   0.896&0.912&0.965 \\ 
\hline $m_A $&12& 12&15 & 2& 2&3\\ 
\hline 
\end{tabular}
   \caption{Time regimes: Summary of  data  characteristics  for publications  by HES and DS,  up-dated till  Dec. 12, 2012, in "proceedings": number  of joint publications (NJP) with at least one coauthor (CA);  
   NPmfCA; number of publications with most frequent (mf) CA;
   NP1CA: number of publications with only one CA;
   NDCA: number of different coauthors;
   power law exponent $\alpha$ in Eq. (\ref{eq1}) and $R^2$ fit;   the CA  core measure ($m_A$) \cite{Sofia3a}  is also given}\label{TimingHES}
     \end{table}
\end{document}